\documentclass[aps,prl,twocolumn,groupedaddress,showpacs]{revtex4}
\usepackage{dcolumn}
\usepackage{epsfig}
\usepackage{graphicx}
\usepackage{color}

\begin{document}
\title{What is the stable atomic structure of NiTi austenite?}
\author{Nikolai A. Zarkevich$^{1}$ and Duane D. Johnson$^{1,2}$}
\email{zarkev@ameslab.gov;  ddj@ameslab.gov}

\affiliation{$^{1}$Ames Laboratory, U.S. Department of Energy at Iowa State University, Ames, Iowa 50011-3020} 
\affiliation{$^{2}$Department of Materials Science and Engineering, Iowa State University, Ames, Iowa 50011-2300} 
\date{\today}

\begin{abstract}
Nitinol (NiTi), the most widely used shape-memory alloy, exhibits an austenite phase that has yet to be identified. The usually assumed austenite structure is cubic B2, which has imaginary phonon modes, hence it is unstable.   We suggest a stable austenite structure that ``on average'' has B2 symmetry (observed by X-ray and neutron diffraction), but exhibits \emph{finite} atomic displacements from the ideal B2 sites. The proposed structure has a phonon spectrum that agrees with that from neutron scattering, has diffraction spectra in agreement with XRD, and has an energy relative to the ground state that agrees with calorimetry data.
\end{abstract}
\maketitle

Nitinol (NiTi) is amongst the most used industrial shape-memory alloys \cite{exptNiTi63,exptB65,exptB91}. 
The shape-memory effect versus temperature (T) is associated with a transformation between the high-T austenite and low-T martensite phases. 
The B2 structure (CsCl with $Pm\bar{3}m$ space group no. 221, see Fig.~1a), suggested for the high-T austenite phase by x-ray diffraction (XRD) and neutron scattering experiments on powder \cite{exptB68,exptB72}, is found theoretically to have unstable phonons \cite{B2phononAnomaly,B2phonon1,B2phonon9}. Hence, B2 NiTi cannot be the actual structure. Yet, neutron scattering measurements \cite{Fultz01} find a stable phonon spectrum for the austenite phase.
From single-crystal x-ray diffraction (XRD) and transmission electron microscopy (TEM) experiments, the austenite structure deviates from B2 \cite{exptB65,exptB2unstable71}, especially at lower temperature, and was described as a premartensitic instability \cite{exptB2unstable71}.
So, determining the structure of austenite is paramount to understand NiTi phase stability and transformations, including those associated with shape-memory. 

{\par}Here we predict a stable austenite structure (Fig.~1), which agrees with known data from neutron scattering and calorimetry \cite{Fultz01,exptHsol}, as well as XRD \cite{exptB65,exptB2unstable71,exptB72,exptHsol}. Intriguingly, this structure has \emph{large} atomic displacements from the ideal B2 sites, up to 22\% of the B2 lattice constant at $0~$K -- above that suggested by the Lindemann criterion \cite{Lindemann1910} for melting! This predicted austenite structure cannot be describe by B2 symmetry, although in diffraction it appears B2 on average (Fig.~1b).
Here we describe only one stable representation of the austenite phase, with many other similar states possible, as found for phonon glass behavior (see Fig.~42 in \cite{RevModPhys86p669y2014}), all of which have similar energy per atom and atomic distribution functions.
Increasing the unit cell and considering other representative stable structures does not change our general results.

{\par}From density-functional theory (DFT), the NiTi ground-state structure is base-centred orthorhombic (BCO) \cite{nmat2p307y2003,GudaVishnu2010745}, while the experimentally assessed structure is monoclinic B$19'$, a low-energy deformation of BCO groundstate \cite{exptB65,nmat2p307y2003}.
In DFT, the previously considered ideal $B2 \leftrightarrow BCO$
transformations \cite{B2ddj9,inprep2014}  have no barrier. 
For ideal B2, the XRD pattern actually differs from those measured \cite{exptB65,exptB2unstable71,exptHsol}.  
In addition, from previous and present DFT results, the B2 energy relative to BCO is 48 meV/atom (557~K), much too high compared to the observed $T_c \approx 313\,$K \cite{exptHsol}. 
Because the ideal B2 phase has multiple unstable phonon modes (Fig.~2), one can expect multiple nearby local energy minima, all with similar energies. 
Lastly, single-crystal diffraction data suggests evidence that NiTi austenite structure was not a simple B2, as believed based on powder diffraction; rather it has a $9.03\,${\AA} superlattice and a $\sim3\,${\AA} sublattice \cite{exptB65,exptB2unstable71}, i.e., a $3\times3\times3$  set of 2-atom B2 cells, containing 54 atoms. This structure was not identified.

\begin{figure}[t]
\includegraphics[width=80mm]{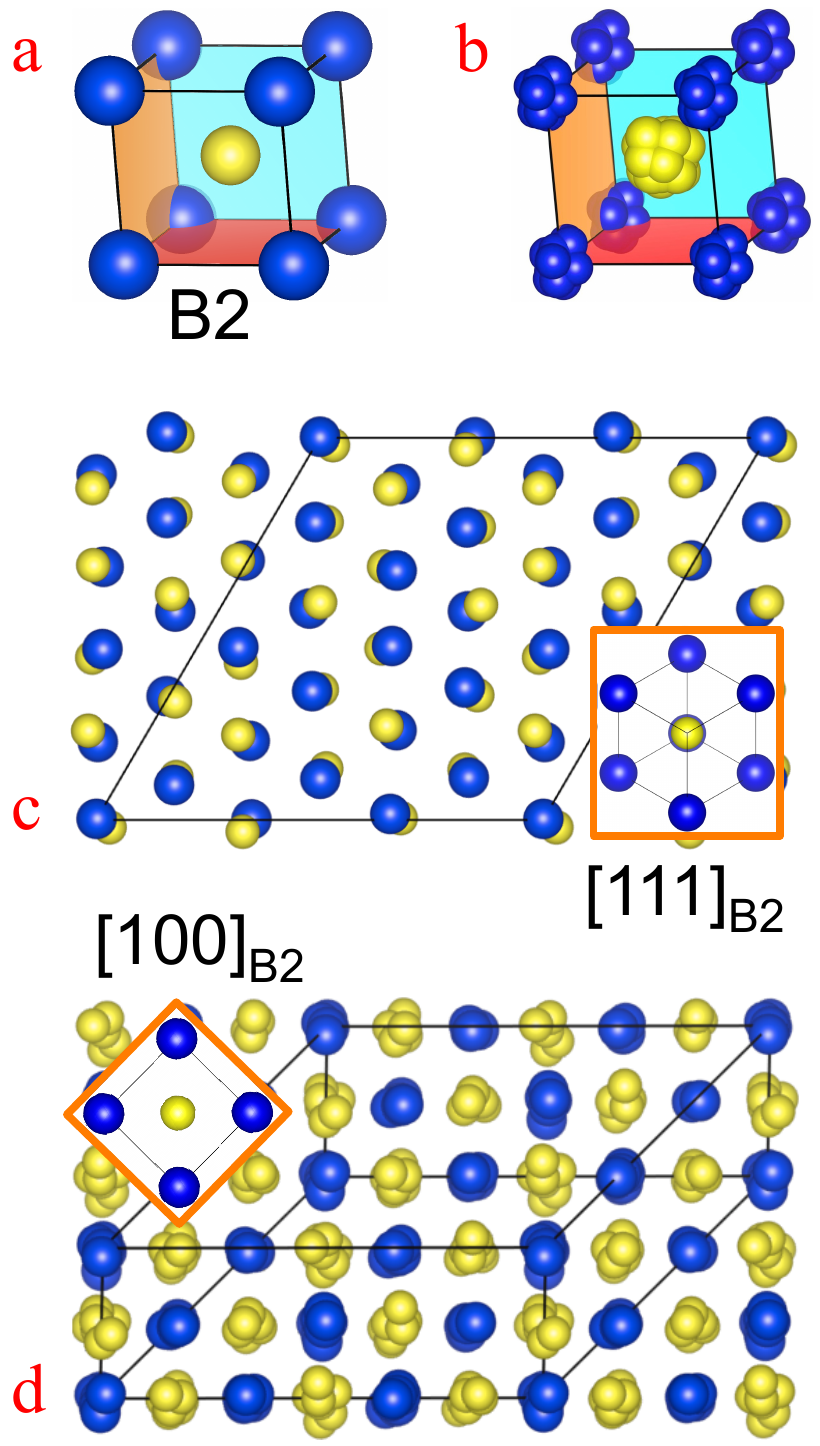}
\caption{\label{fig1_str}
Atomic structure of the Ni$_{27}$Ti$_{27}$ austenite, compared to B2, with Ni (yellow) and Ti (blue) atoms. 
(a) Ideal B2.
(b) Projection of Ni$_{27}$Ti$_{27}$ atomic positions onto a B2 cell. 
(c) Hexagonal [0001] projection, compared to [111]$_{\mbox{B2}}$.
(d) Viewed along [100]$_{\mbox{B2}}$.   }
\end{figure}

{\par } To find a stable NiTi austenite, we investigated cells of increasing size with B2 chemical order that permitted symmetry breaking by atomic displacements. 
We chose a 54-atom hexagonal unit cell with the $c$-axis along B2 $\left<111\right>$, and two $\sim$$9\,${\AA} basal plane vectors along B2 $\left<\bar{1}01\right>$ and $\left<0\bar{1}1\right>$.
Its unit cell vectors $(\bar{a},\bar{b},\bar{c})$ in terms of $(\bar{x},\bar{y},\bar{z})$ for cubic B2 are 
$\bar{a}=3(\bar{z}-\bar{x})$, $\bar{b}=3(\bar{z}-\bar{y})$, and $\bar{c}=\bar{x}+\bar{y}+\bar{z}$. 
Using \emph{ab initio} molecular dynamics (MD) followed by relaxation at 0 K to a local energy minimum, we obtain a stable austenite Ni$_{27}$Ti$_{27}$ structure (Figs.~\ref{fig1_str}c-d). 
This hexagonal structure (Fig.~1) has a $c/a$ of 0.3954, which is a 3\% 
reduction compared to $c/a$ of 0.4082 (i.e., $1/\sqrt{6}$) for our unit cell with ideal B2 order, 
where $c=a_{B2}\sqrt{3}$ and $a= 3a_{B2}\sqrt{2}$  in terms of $a_{B2}=3\,$\AA.”

{\par } Our austenite structure has a DFT energy $\Delta E$ of $29~m$eV/atom ($340~$K) above BCO, in better agreement \cite{dEkT} with the measured $T_c$ for equiatomic NiTi of $313\,$K \cite{exptHsol}.  
Typically, $\Delta E \approx k_B T_c$ \cite{dEkT}, which provides reliable estimates of phase transition temperatures \cite{CoPt2010}, especially for magnetic transitions.
From the $\Delta E$ and estimated entropy difference of 1/2 $k_B$ per atom \cite{Fultz01} for the martensite-to-austenite transformation, we predict \cite{PRL100p040602} the latent heat to be below 15 meV/atom (1.4 kJ/mol) at $T_c$, while the calorimetry values are $1.07 \pm 0.10$ (cooling) and $1.3 \pm 0.2 $ kJ/mol (heating) \cite{Fultz01}.  Thus, the calculated thermodynamic quantities closely reproduce the measured values.

\begin{figure}[t]
\includegraphics[scale=0.36]{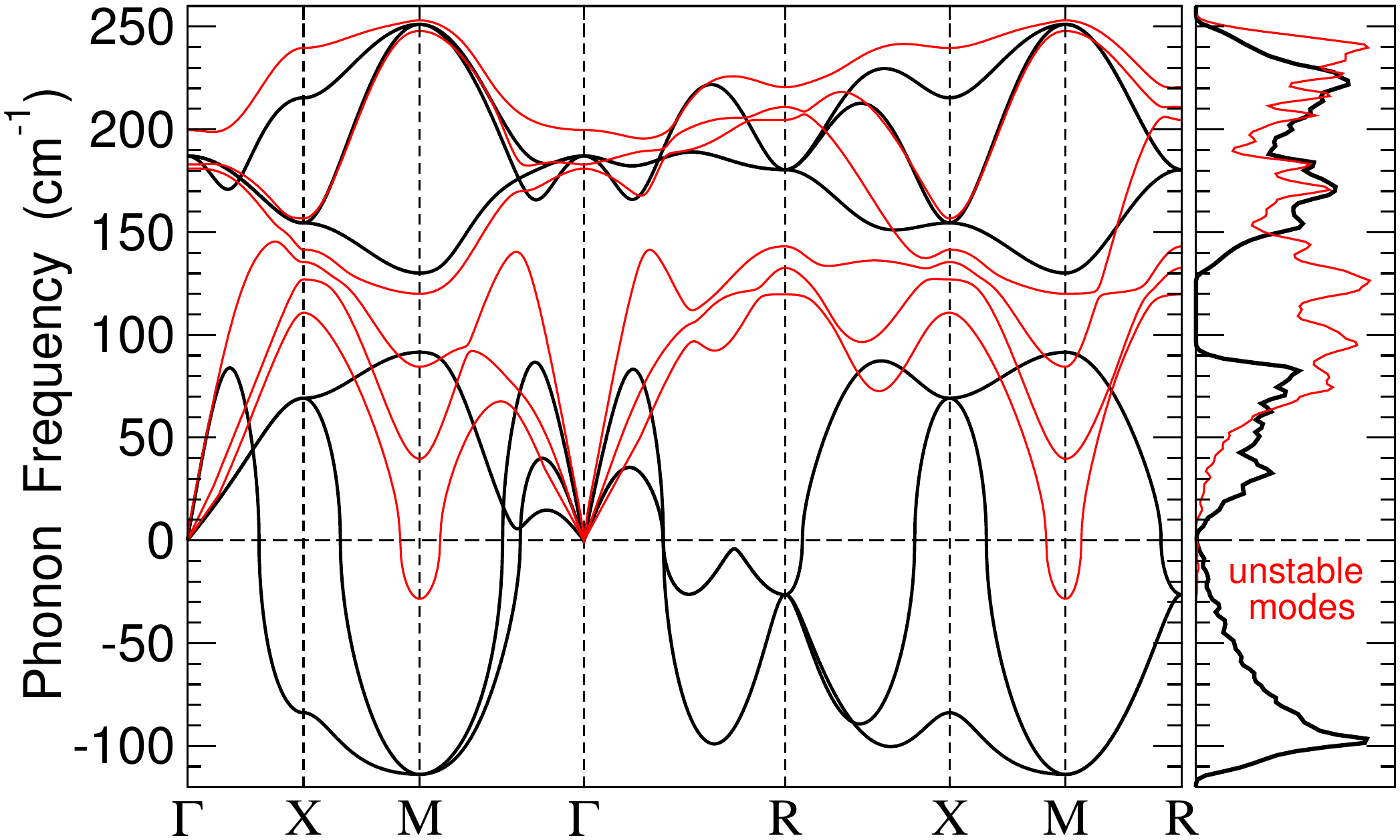}
\includegraphics[scale=0.36]{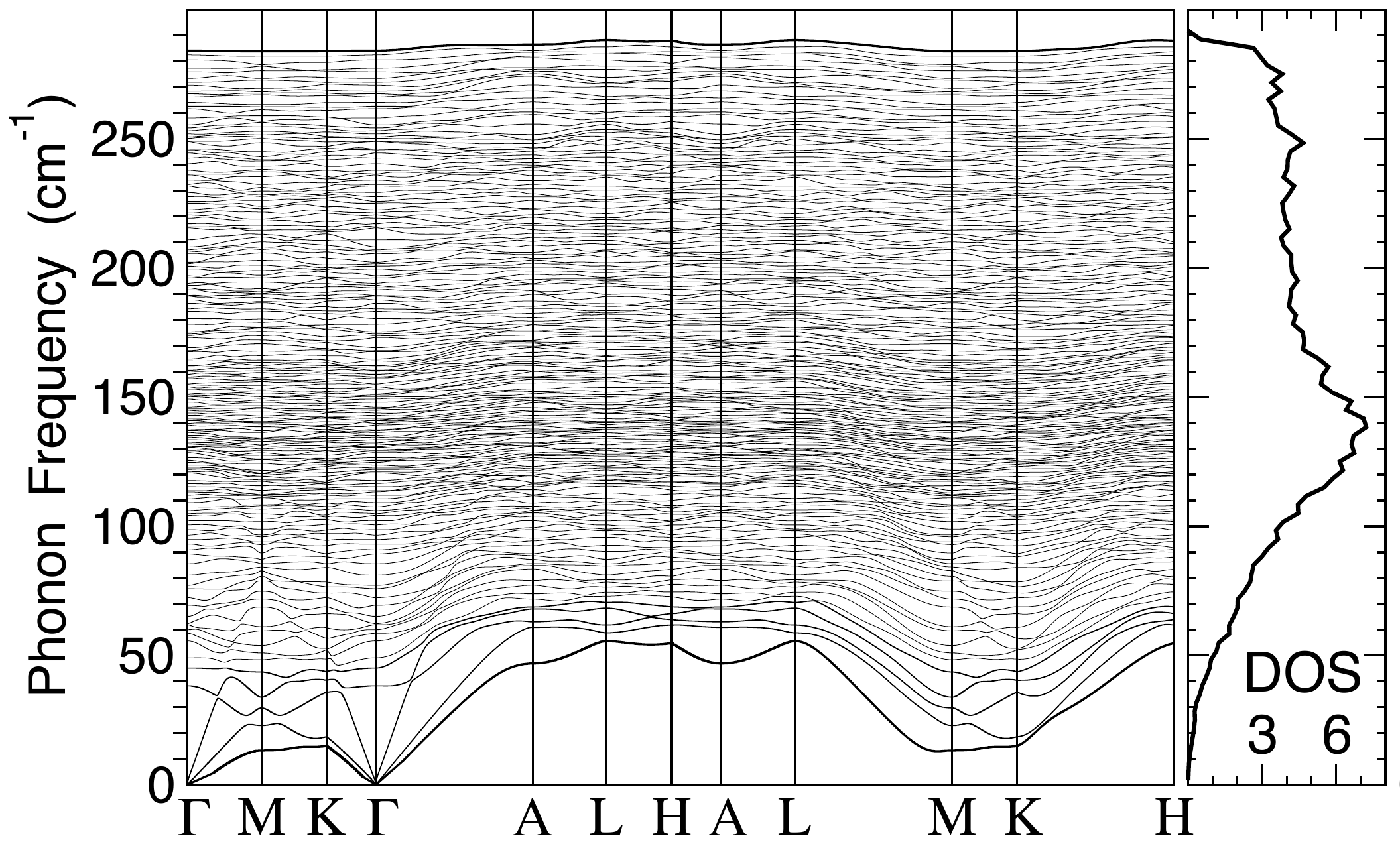}
\caption{\label{fig2phonons}
Calculated phonon spectra and density of states (DOS) of (upper) ideal B2 at $0\,$K (black) and $1586\,$K (red) with unstable  negative phonon frequencies and (lower) new austenite structure (stable), with DOS units of $10^{-3}$ states/cm$^{-1}$.}
\end{figure}

\begin{figure}
\includegraphics[width=80mm]{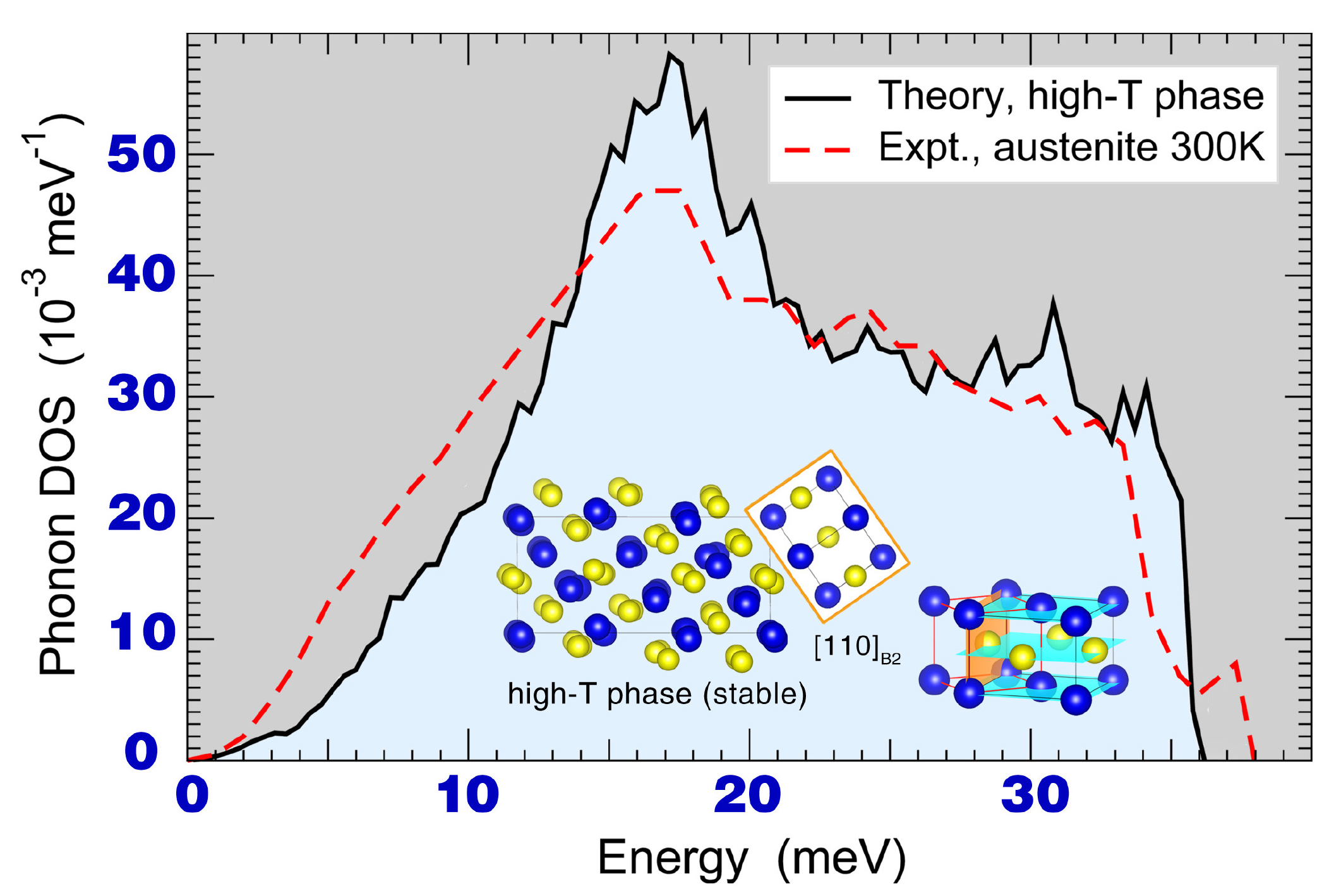}
\caption{\label{fig3DOS}
Austenite phonon DOS from theory (new structure, see Fig.~2) and neutron diffraction experiment \cite{Fultz01}.
Experimental error bar below $10~me$V is $\pm 4~me$V. Insert: austenite structure (left) compared to ideal B2 (right) and its [110]$_{\mbox{B2}}$ projection (center).}
\end{figure}

{\par }\emph{Computational Details:} To predict this structure, we used a plane-wave psuedo-potential-based DFT method  using the generalized gradient approximation (GGA) \cite{GGA} and a projected augmented wave (PAW) basis \cite{PAW}, as implemented in VASP code \cite{VASP1,VASP2} with convergence obtained by a second Broyden's method \cite{BroydenDDJ}. 
We choose 337~eV plane-wave energy cutoff and $544.6~e$V augmentation charge cutoff. 
We converged total energies and forces using $k$-meshes with at least 50 $k$-points per {\AA}$^{-1}$~(e.g., $11 \times 13 \times 17$ for a $4.92 \times 4.00 \times 2.92$ {\AA} cell). 
The structure of the high-T phase in a 54-atom Ni$_{27}$Ti$_{27}$ cell is investigated using \emph{ab initio} MD with 1 fs time steps in a Nos\'e thermostat. After 1000 fs at $800\,$K, temperature was quenched from $800\,$K to $0\,$K in 800 fs; next, the atoms were relaxed using the conjugate gradient algorithm. We performed an internal atomic relaxation in a large fixed unit cell, and then a full relaxation of both atoms and lattice vectors. 
Lastly, phonon spectra were constructed using the calculated atomic forces for 162 independent 0.04~{\AA} displacements 
in a 108-atom 1$ \times $1$ \times $2 hexagonal supercell (only 2 independent atomic displacements in a 54-atom 3$ \times $3$ \times $3 cubic supercell for B2) within the small displacement method in the PHON code \cite{Phon}. We have checked that the Phonopy \cite{Phonopy} code gives similar results.

{\par }
Phonons at finite T (Fig.~2) are addressed by combining 3 codes: VASP \cite{VASP1,VASP2}, ThermoPhonon \cite{ThermoPhonon}, and Phonopy \cite{Phonopy}. First, an \emph{ab initio} MD in a 54-atom, 3$ \times $3$ \times $3, $(9\,\mbox{\AA})^3$ cubic supercell with Nos\'e thermostat at a given T with 1~fs time steps is used to obtain atomic positions and forces for over 50000~fs steps after 2000~fs equilibration. Next, force constants are calculated using our ThermoPhonon \cite{ThermoPhonon} code in the assumption of ideal B2 average atomic positions, 
and used in Phonopy \cite{Phonopy} with symmetrization to construct the phonon spectrum from the \emph{ab initio} MD data.
We performed MD calculations at a range of T, including 0, 300, 800, 1200, and $1586\,$K (melting), and found that although already at $300\,$K the instability is only  around $M$, and its relative weight decreases with T, this instability survives at all T up to melting. 

\begin{figure}[t]
\includegraphics[width=80mm]{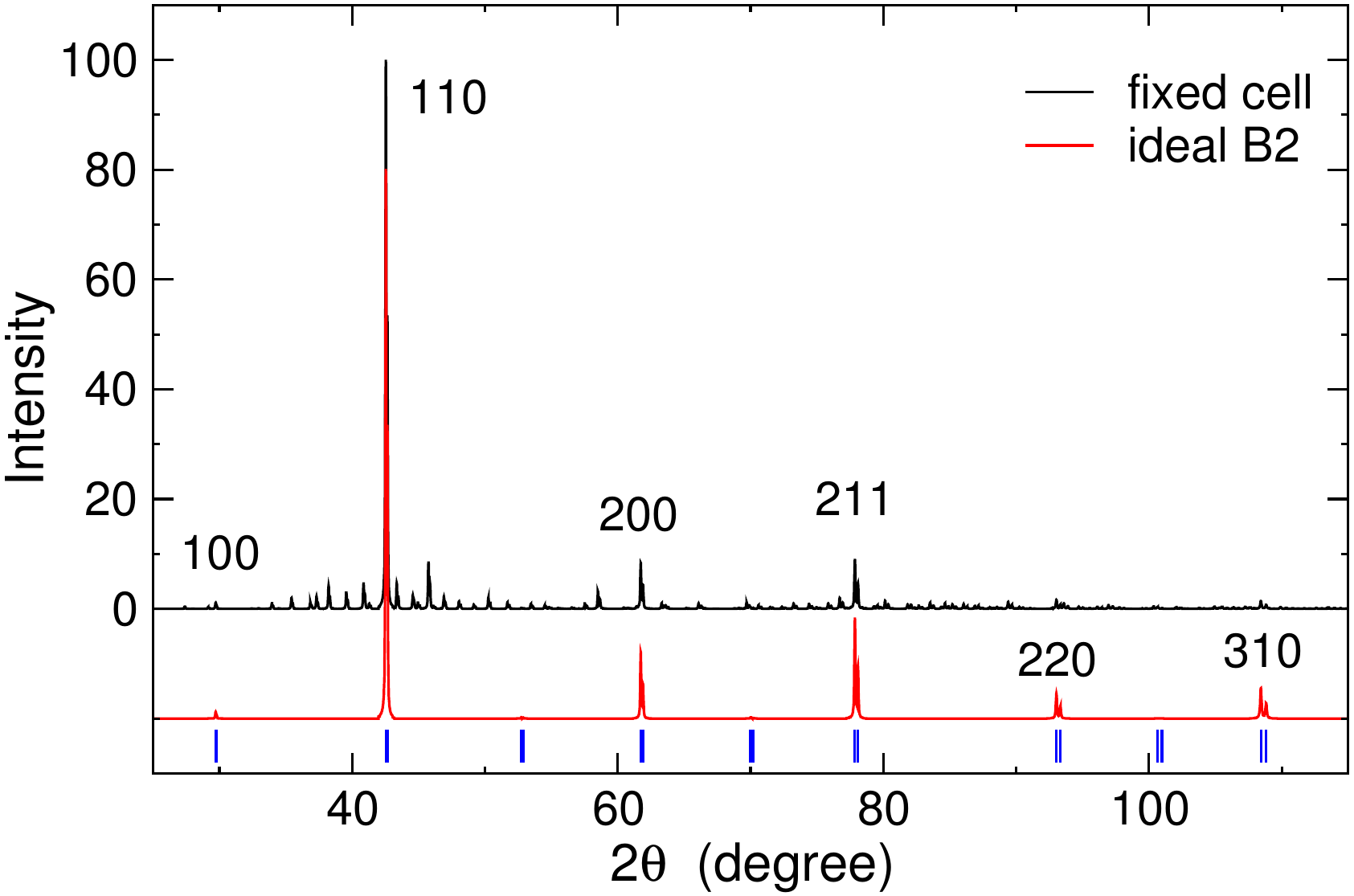}
\caption{\label{fig4XRD}
Simulated XRD spectrum of our proposed austenite structure (black) compared to ideal B2 (red).}
\end{figure}

{\par }For the proposed austenite structure, the calculated $0\,$K phonon spectrum is stable (Fig.~2), in contrast to ideal B2.
Also, as we have confirmed up to $1586~$K (experimental melting temperature), vibrational entropy does not stabilize ideal B2, see Fig.~2, where the $M$ point remains always unstable at high temperatures. 
More compellingly, the calculated phonon density of states (DOS) of the predicted austenite phase (Fig.~2) agrees well with that found from neutron scattering \cite{Fultz01}, as shown directly in Fig.~3.  

{\par}Interestingly, this predicted austenite structure, with atoms displaced from the perfect B2 positions, looks like B2 on average (Fig.~1b).
The simulated XRD pattern at $0\,$K is shown in Fig.~4. 
With all B2 peaks still present, this pattern does not contradict any previous experimental XRD data.
The additional XRD peaks can be at various positions for different local energy minima in the austenite phase; in this case they contribute to the background after summation. 

\begin{figure}[t]
\includegraphics[width=75mm]{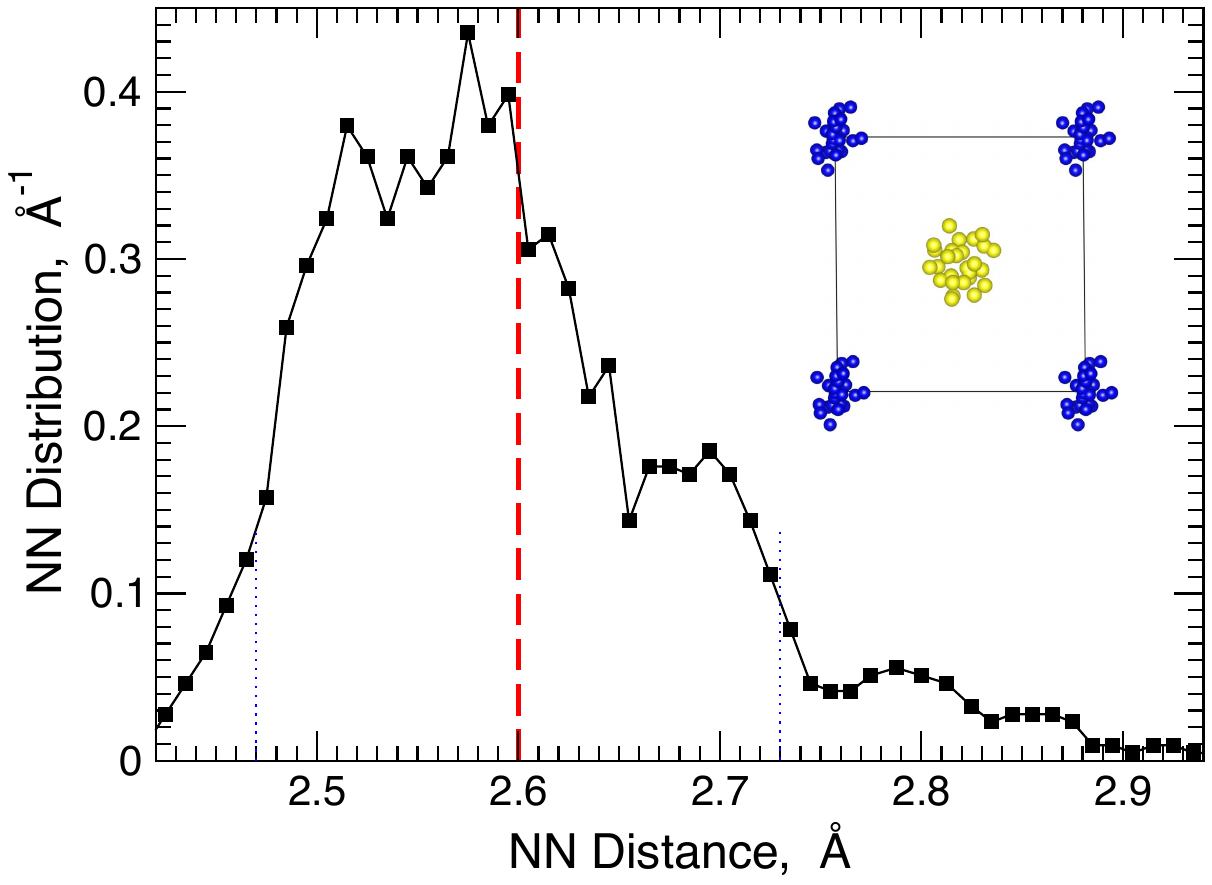}
\caption{\label{Fig5_NN_Hist}
The NN-pair distribution function in the fully relaxed austenite structure (atomic coordinates are in Table S1). The binning size is $0.04\,${\AA}. The NN distance in ideal B2 structure is $2.6\,${\AA} (vertical dashed line); dotted lines show $\pm 5$\% deviations. Insert: Ni (yellow) and Ti (blue) positions projected onto a B2 cell (see Fig 1b).}
\end{figure}

\begin{figure}[]
\includegraphics[width=80mm]{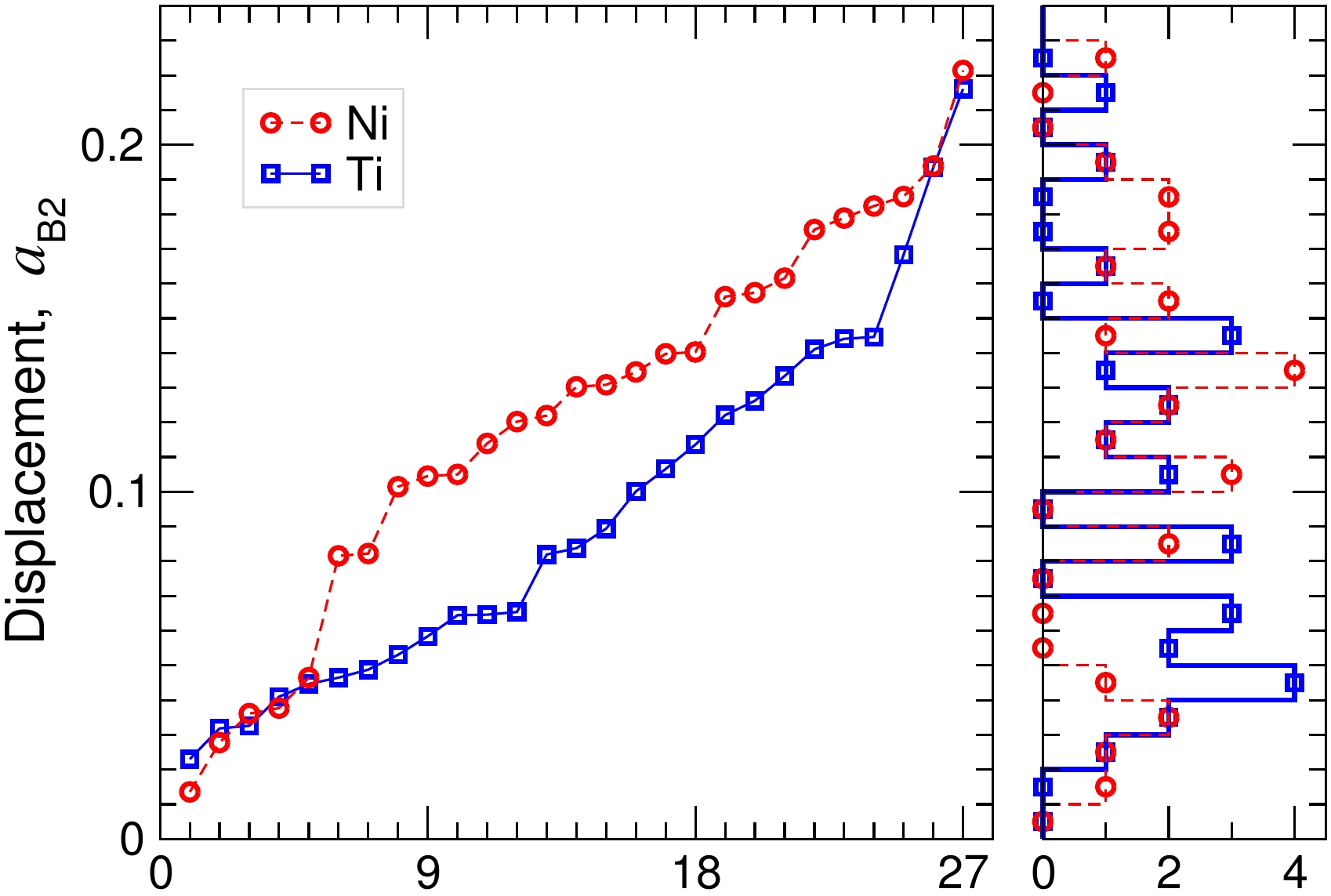}
\caption{\label{Fig6_Site_Displ}
Displacements from ideal B2 positions (sorted by magnitude) of the relaxed atoms in a fixed hexagonal Ni$_{27}$Ti$_{27}$ unit cell. Right panel: Number of displaced atoms per $0.03\,${\AA} bin, where the largest value ($0.66\,${\AA}) is 25\% of the NN distance and 22\% of the lattice constant in ideal B2, $a_{B2}=3\,$\AA.}
\end{figure}

{\par} Further details of the structure offer additional insight, and potential comparison to pair distribution functions from new diffraction experiments. From the calculated pair distribution function, the nearest-neighbor (NN) distances have a distribution (Fig.~\ref{Fig5_NN_Hist}) with $<$5\% half-width from the $2.6\,${\AA} NN distance in B2 
($2.43$ to $2.88\,${\AA}). 
In the supplement, we provide the atomic direct lattice coordinates (Table S1) in our representation of the austenite unit cell. From this data, we plot the NN-pair distribution function (Fig.~\ref{Fig5_NN_Hist}) and the atomic displacements relative the ideal B2 positions (Fig.~\ref{Fig6_Site_Displ}). 

{\par } 
In summary, there is a multiplicity of solutions to our original question: 
What is the stable structure of NiTi austenite?
We have proposed a stable representative structure (one of many) for the high-T NiTi austenite phase, whose energy relative to the BCO ground state, diffraction spectra, and vibrational density of states  agree, respectively, with available calorimetry, XRD, and neutron scattering data, whereas those of an ideal (unstable) B2 do not. We suggest new experiments to assess the inherent displacement in NiTi austenite, with a large Debye-Waller factor at low temperatures (not the usual thermal disordering from phonons).  We are also assessing the solid-solid martensitic transformation paths for NiTi \cite{inprep2014}.

\acknowledgements
We thank Dario Alf\'e, Brent Fultz, and Graeme Henkelman for helpful discussions. 
This work was supported by the U.S. Department of Energy, Office of Science, Basic Energy Sciences, Materials Science and Engineering Division. The research was performed at the Ames Laboratory, operated for the U.S. Department of Energy by Iowa State University under contract DE-AC02-07CH11358.

\end{document}